# USER PROFILE RELATIONSHIPS USING STRING SIMILARITY METRICS IN SOCIAL NETWORKS

VASAVI AKHILA DABEERU

## Abstract:

This article reviews the problem of degree of closeness and interaction level in a social network by ranking users based on similarity score. This similarity is measured on the basis of social, geographic, educational, professional, shared interests, pages liked, mutual interested groups or communities and mutual friends. The technique addresses the problem of matching user profiles in its globality by providing a suitable matching framework able to consider all profiles' attributes and finding the similarity by new ways of string metrics. It is able to discover the biggest possible number of profiles that are similar to the target user profile, which the existing techniques are unable to detect. Attributes were assigned weights manually; string and semantic similarity metrics were used to compare attributes values thus predicting the most similar profiles. Profile based similarity show the exact relationship between users and this similarity between user profiles reflects closeness and interaction between users.

## 1. INTRODUCTION

Internet technology has mobilized people around the world to re-gestate their image and dramatically change the landscape of its identity construction. It has become the ultimate platform for accelerating the flow of information with various social networking sites. The transformation in community from densely knit villages and neighborhoods to more sparsely knit social networks was fostered by the internet and thus became fastest-growing form of social media which changed the way of communication among people.

The Internet provides the capacity to mine data on the behavior of users as well as to disseminate information. Some general characteristics of social networks are given in [1]-[12]. A social network may be viewed as a directed graph where each user is a node. The relationship between interacting cliques may be studied using the theory of feedback networks [13]-[16]. The structural properties of social networks include scale invariance and the small world phenomenon [17]-[21]. For examining certain kind of interaction between nodes one may use instantaneously trained networks [22]-[28]. Prediction using social media is discussed in [29]-[31]. Social network dynamics based on game theory are discussed in [32]-[36]. In aggregation of nodes it is essential to classify them using an appropriate metric, which is the study of the present article. Once the aggregation is done then the task of the analysis of the graph representing the network as well the problem of determining structural relationships using complex system theory becomes easier.

Online social networking websites became a platform for users to express themselves beyond physical features and labels, interacting with each other to share experiences, discuss interests, and influence one another in a selective network [9]. It helps in allowing users to make and develop social relationships with individuals of similar interests around the world. That means, they are not constrained by the same geographic boundaries and which it became an enduring part of everyday life.

According to a survey that carried out on internet users, Facebook was marked as the most popular social networking website all over the world counting to more than 500 billion members. It is noted that each user spends 30 minutes on average per day on the site and so became an important phenomenon on the Internet [29]. Social networking sites have had much effect on the way users maintain relationships, on their number and their diversity [3]. Friends tend to come from similar social-demographic backgrounds, share common interest and information.

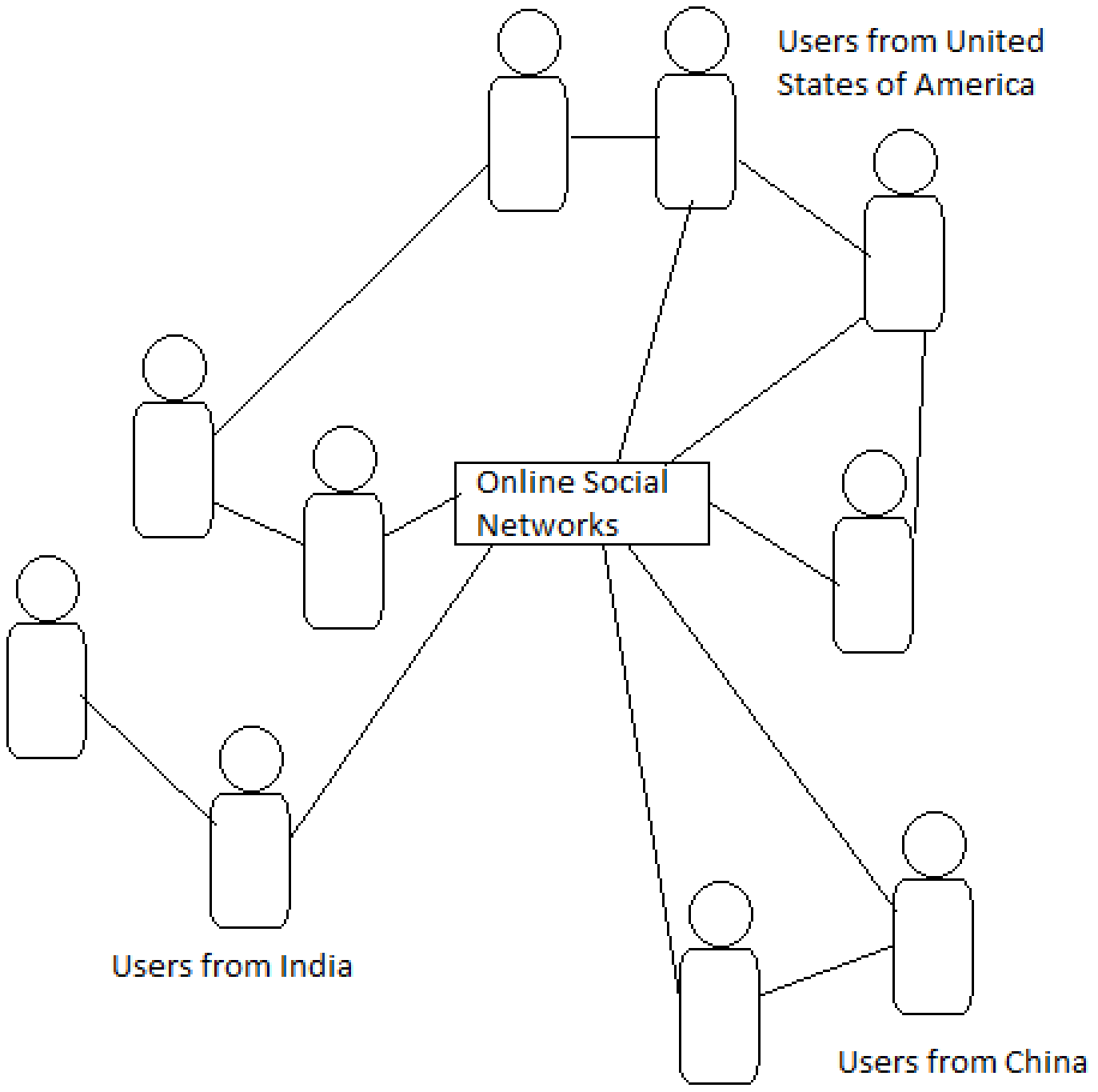

**Figure 1 Interaction of various users around the world using the online social networks**

Facebook, the most popular social networking website, allows users to create personal profiles viewable to anyone in a given network. Individuals can enter information on their background (e.g School, Hometown, Study ), interests, political views, demographics( e.g Birthday, Gender ) , cultural views ( e.g Religion), favorites such as books, movies, shows, music etc. Additionally, users can enter “friendship” relationships with other registered users and share photo albums that can be linked to the profiles of those present in the picture [21]. The assumption in this characterization is that users are completely defined by their self-identified characteristics.

The remaining section is organized as follows: section 2 includes the background work on Facebook social network; section 3 lists graphical representation of Facebook social network to describe how users are connected to each other; section 4 applies to the proposed architecture of theoretical model which describes how similarity score is calculated and shows the interaction or closeness level between the users.

## 2. BACKGROUND

Facebook is a social utility that helps people understand the world around them. People with a valid email address can register for Facebook and create a profile to share information with their friends in a trusted environment. This new tool helped the individuals to connect and most importantly communicate with friends and other users in the network [5] [10]. Each user has an individual profile where their personal and professional information can be shared. An individual user home page consists of many features such as the ones given below.

**News Feed**: This customizable version of the profile highlights information that includes profile changes, upcoming events, and birthdays, among other updates. It also shows conversations taking place between the walls of a user's friends.

**Timeline**: A Timeline is the new virtual space in which all the content of Facebook users is organized and shown. The photos, videos, and posts of any given user are categorized according to the period of time in which they were uploaded or created.

**Notifications**: Notifications are what inform the user that an addition has been added to his or her user profile page.

**Groups**: Facebook Groups can be created by individual users. It allows members to post content such as links, media, questions, events, editable documents, and comments on these items and security constraint can be added to the groups by setting the privacy settings to open, closed or secret.

**Like**: It is described by Facebook as a way to "give positive feedback and connect with things you care about", users can "like" status updates, comments, photos, and links posted by their friends, as well as adverts.

Some other features can be viewed in the below given table.

**Table 1 Site Features of Facebook Network**

| Site Features |
| --- |
| Pages and Communities |
| Messages and inbox |
| Notifications |
| Events |
| Photos and Videos |
| Place and status |

### 2.1 User Profile to Friends Networks

The friend’s network of Facebook, our topic of study, has two varieties of accounts: users and communities/groups [17].

**Table 2 User, Group, Link Relationship**

| Source Point | Destination Point | Link Description |
|---|---|---|
| User | User | Friendship or trust |
| User | Group/Community | Subscribership or Readership |
| Group/Community | User | Membership ,posting access |
| Group/Community | Group/Community | Outmoded |

The above Table 2 shows the types of links in Facebook and their constituent attributes. Friendship is an asymmetric relation between two accounts, each represented by a vertex in a directed graph. For example, when a user x adds another user y to his or her friends list, he/she can specify the membership in any of the groups or communities.

**Friends**: A user has a connection with their friends by adding them to their friends network. The Facebook gives the user an opportunity to add them to their close friends, acquaintances, family or by creating a new list.

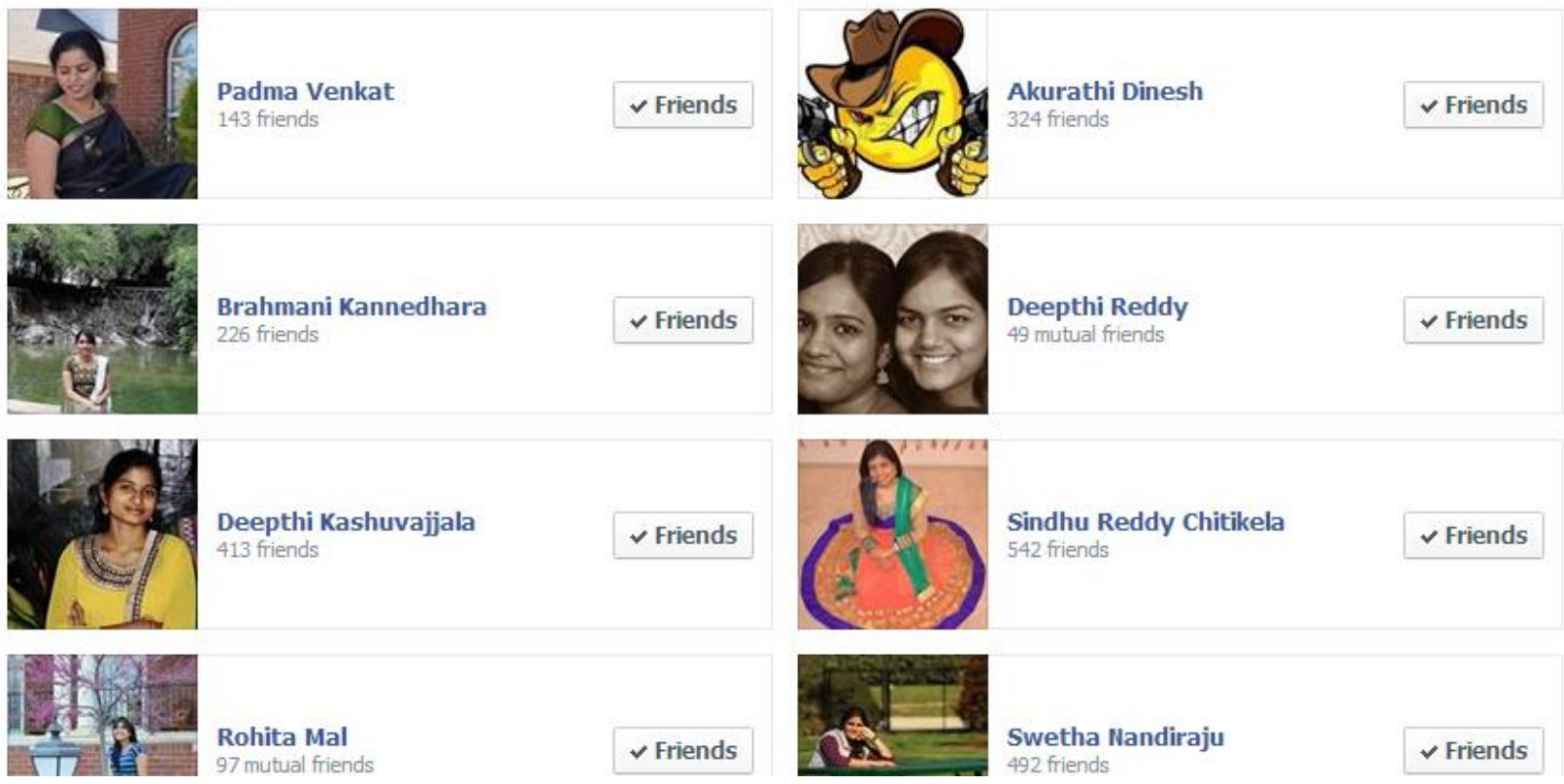

**Figure 2 Facebook Friends List**

**Groups / Communities**: Facebook Groups are dedicated to group discussion on topics of common interests. Facebook group pages can be made public, where anyone can join, closed or secret.

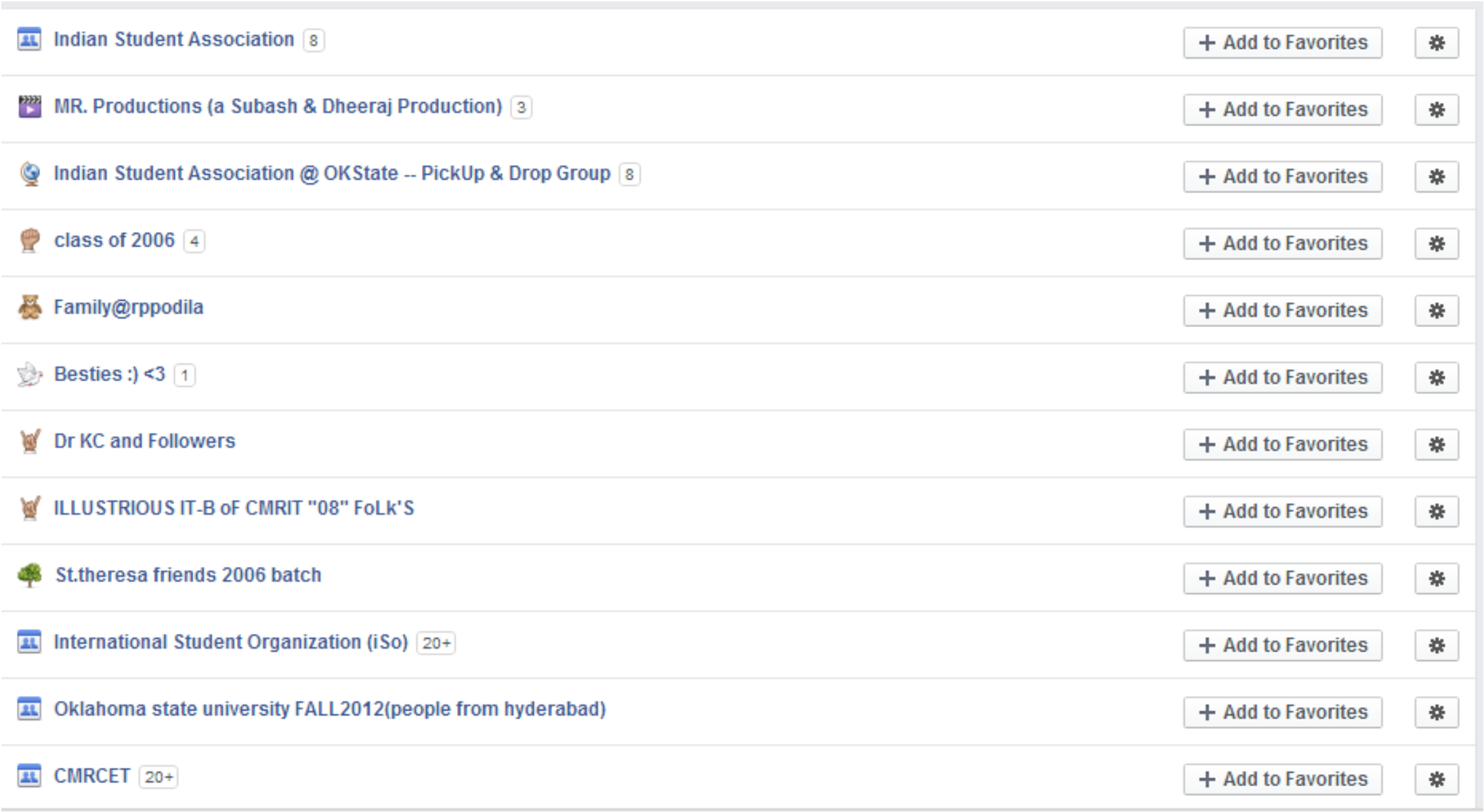

**Figure 3 Facebook Groups / Communities**

### 2.1.1 Identifying connection among users

Users in online social network visualize as a node and the link between users reflects the relationship. Our initial approach to link identification consisted of dividing friend's network features into graph features [1].

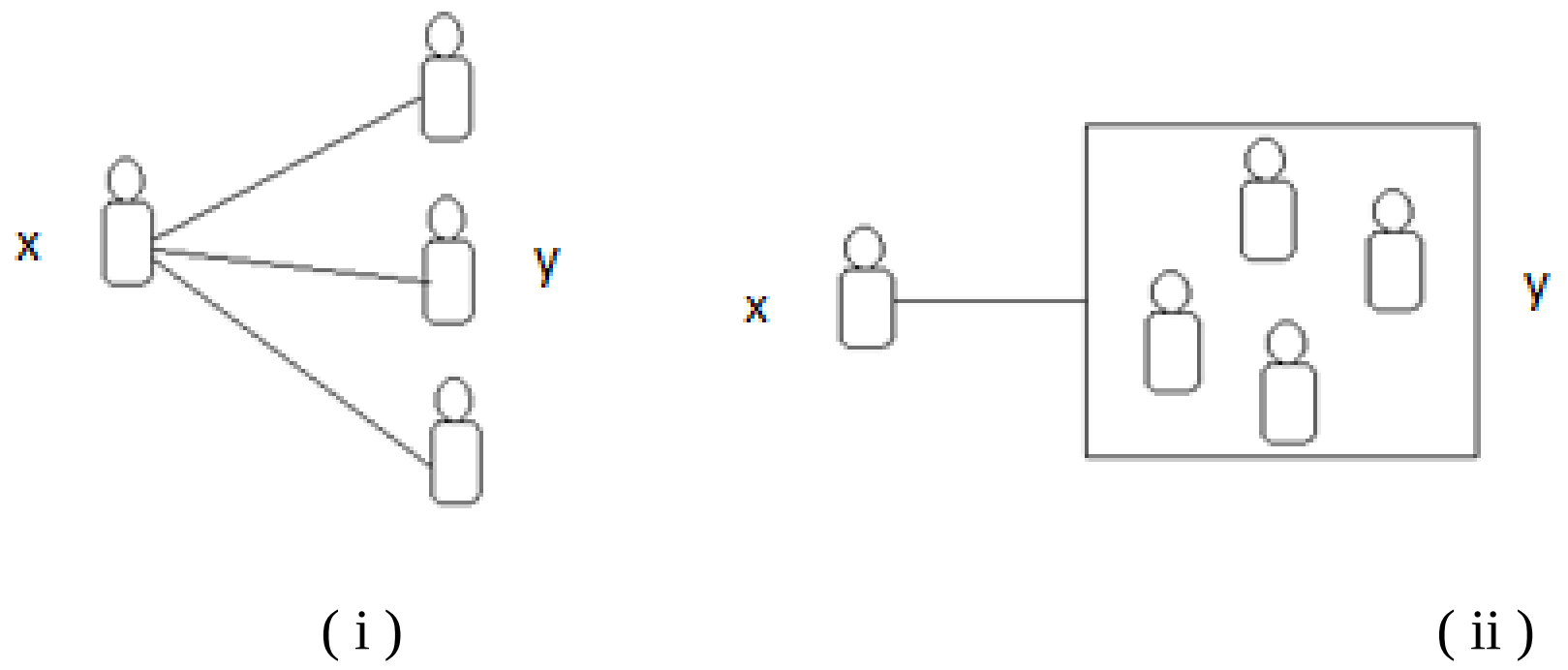

**Figure 4 ( i ) User to User/Users Communication ( ii ) User to Group Communication**

In the above figure 4, it depicts the user to user communication and user to group. The user to user communication can be one to one or one to many to relationship. The user to group communication can be one to one or one to many and many to one relationship. The concept of social group or community is to build up a community that is based upon an interest, same view, likeness or dislikeness or some kind of association. Thus classification and grouping of users is a pattern classification problem for which different techniques exist. These include a variety of techniques based on neural networks [13]-[17], [24]-[27]. Sometimes the patterns can be efficiently represented by graphical techniques as shown in the next section.

### 2.2 Graphical Representation

In network analysis, one is interested in the connections between users. These connections can be easily represented by graphical representation of a network. Graphical representation of the network is generated by a java applet, called Gephi that lays out the largest connected component of the graph. It is an interactive visualization and exploration platform for all kinds of networks and complex systems, dynamic and hierarchal graphs [6]. It helps to explore and understand graphs easily. This applet interface allows one to locate specific users within the global graph [11].

#### 2.2.1 Information Retrieval for visualization of Facebook Network

When an individual logs in his/her Facebook account, a web data extractor which is present in Facebook ,called netvizz can be used to retrieve the information of a particular individual. The web data extractor retrieves the user friend’s network connections as a gdf or .net extension file. This file is given as input to Gephi, for graph visualization. It takes the input file and produces a graph with nodes and edges connecting it. Initially the graph consists of nodes with no labels.

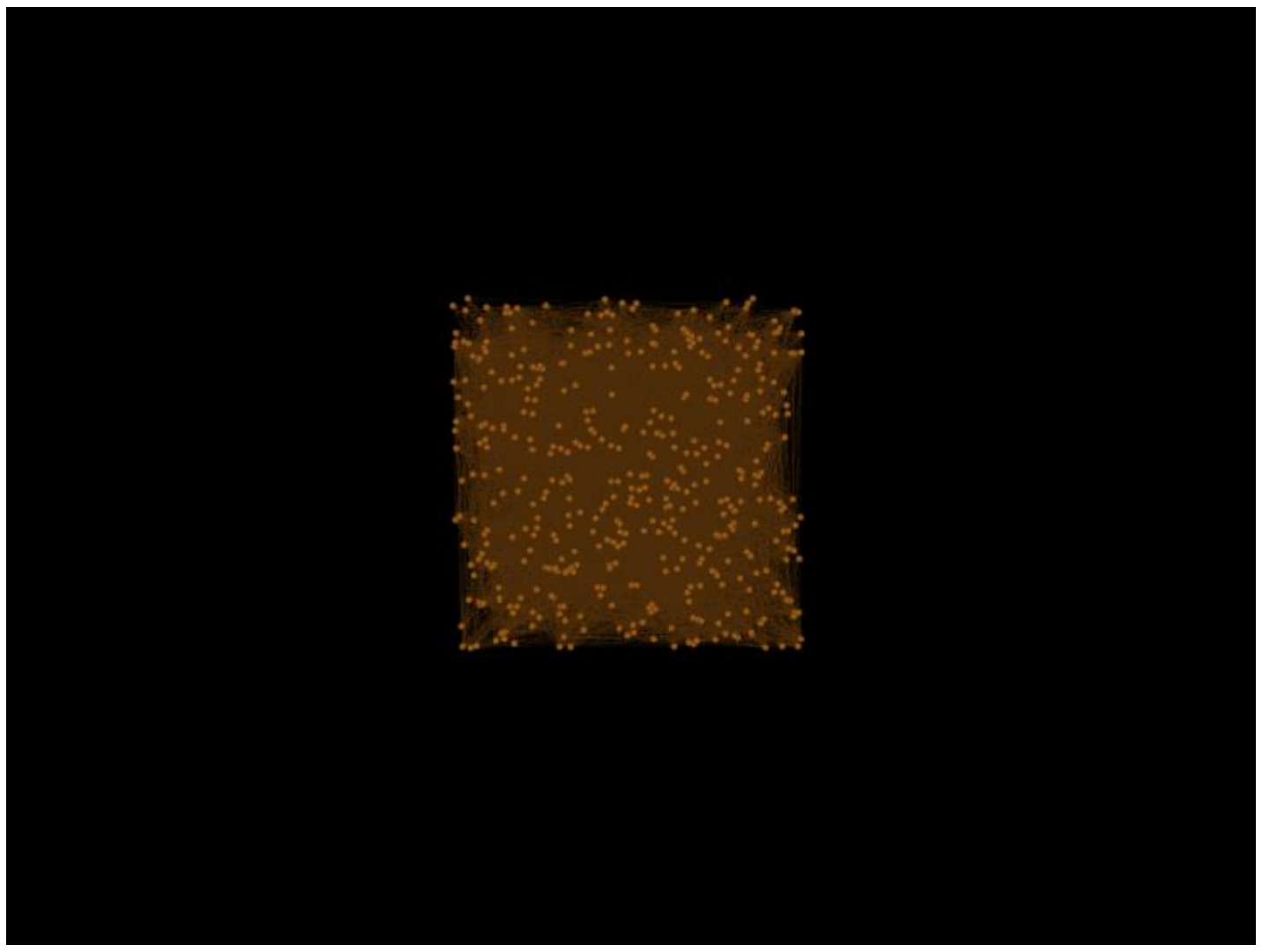

**Figure 5 Facebook User Network Graph**

This is the preview graph where the edges are arcs instead of straight lines. The node names are also displayed The blue arcs are the connectors and orange nodes represent the users.

Figures 5 and 6 are snapshots of the Facebook profile users. Fig 5 shows the network graph of Facebook user and his/her connections with other users. Each user is treated as a node/ vertex in a network graph and their relation with other nodes or vertices as an edge. Fig 6 represents the Facebook network of an individual with nodes containing node labels and its connection with other profiles. The graph is represented as a force-based atlas layout, where nodes attracted are closely knit together, as similar profiles. Each layout has its own implementation.

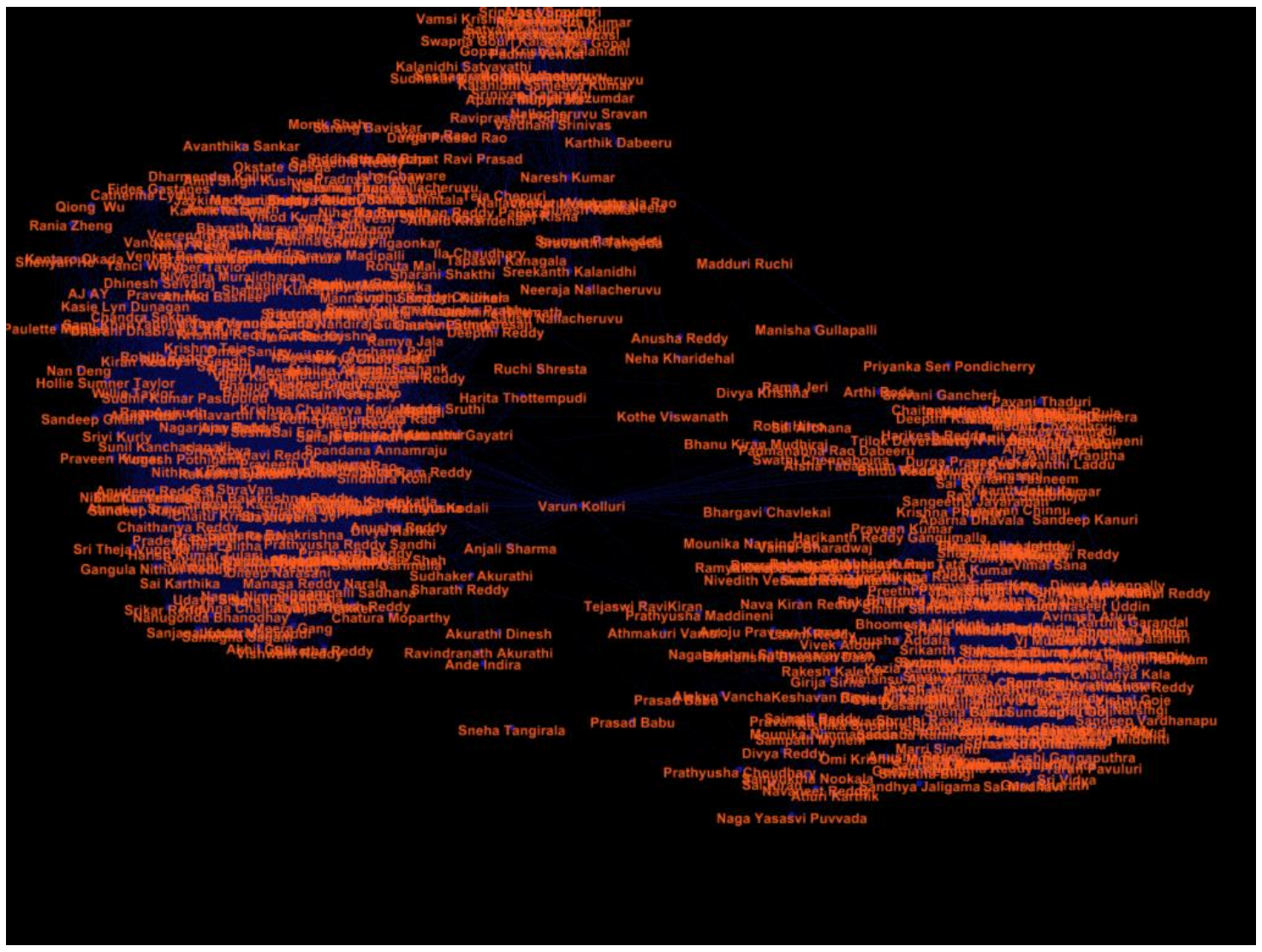

**Figure 6 Facebook User Network Graph with node labels**

Graph visualization using Gephi allows for a bigger picture, combining several key elements of the network into an interactive graph with the potential for the users to manipulate the structures and colors to review hidden properties and patterns. It allows users to check statistics by a data table. Each graph consists an excel sheet of input data that includes the information about the all nodes such as NodeId, NodeLabel, Agerank etc

## 3. SIMILARITY METRICS

### 3.1 Introduction

Of all the techniques used in predicting the similarity between profiles, we use string similarity metrics and find a new similarity score between profiles which is easy and simple in implementation though the analysis is done using large network of Facebook users. There are various ways in predicting the similarity of profiles, but the technique used, allows the users to give importance to some attributes and assign weights as well as compare each string with the other gives more efficient and reliable way of detecting the similar profiles.

#### 3.1.1 Identification Connection between Two User Profiles

A user has connection with his/her friends forming a network by maintaining a friends list which consists of links of all friends' profile [4]. Users in the online social network are represented as a node or vertex and the link between users or groups represents the edge of the network [8].

#### 3.1.2 Exploration of User's Profile

Investigate the relationship between two or more profiles by crawling from source profile to destination profile so as to check the closeness or interaction level between source end and destination end [5].

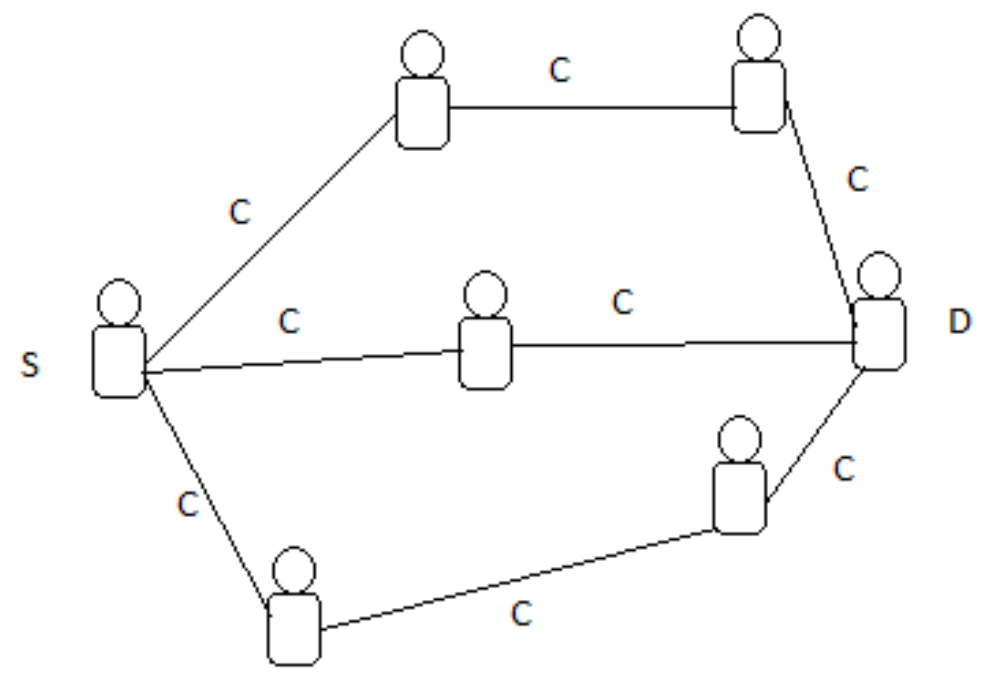

**Figure 7 Interaction / Closeness Level of Users**

C is the closeness level from source point ( S ) to destination point ( D )

### 3.1.3 Identification Of Closeness Or Interaction Level

A strong connection level with other profiles can be extracted and estimated. The closeness or the interaction level between two profiles can be calculated based on the following factors.

1. On the basis of communication amongst the profiles
2. On the basis of profile similarity

In an online social network, the users are connected with each other based on various aspects. The link between them doesn't depend on the distance among the users. Each user is considered as in virtual world and individuals are not considered by the geographical distance; instead it just means the psychological distance and is measured by the influence maintained among the users of the network [5].

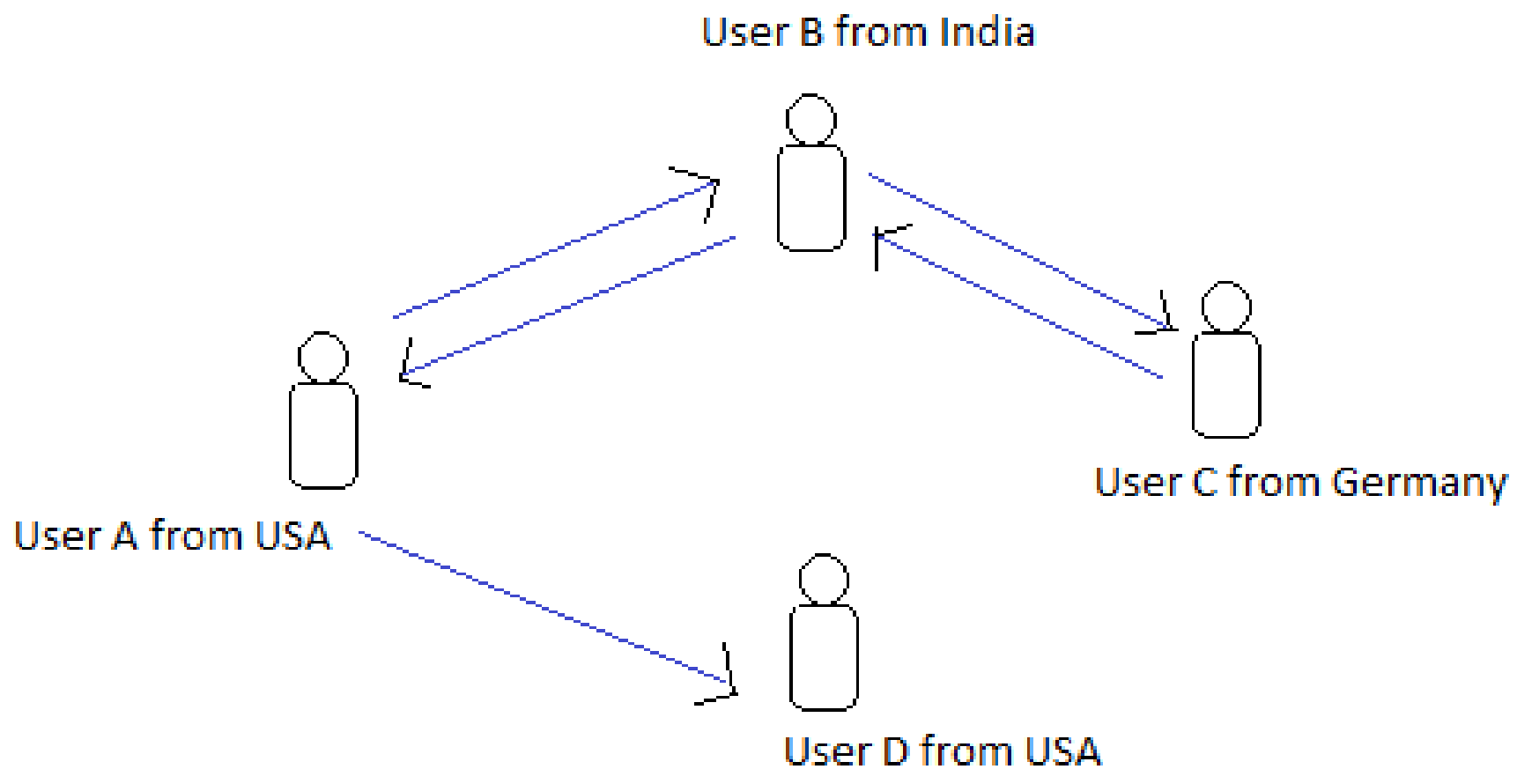

**Figure 8 Communication among users considering distance between them as psychological distance**

The influence between the individuals also shows the distance between them with respect to contextual similarity since the influence indicates the degree of their shared interest represented as terms. The influence and contextual distance between individuals are inversely related.

When a user creates a Facebook profile, each user has their own user profile page. They can provide their personal, professional and social information. Based on the user profiles created and information provided by the users, closeness of the profiles by compared.

On the basis of contents uniqueness, similarity is measured by analyzing texts, links messages etc. If we are trying to check the closeness that a user U is linked to a user V, we add up the number of items that the users U and V has in common. Hence the items that are unique to few users weight more than commonly occurrence of items [4].

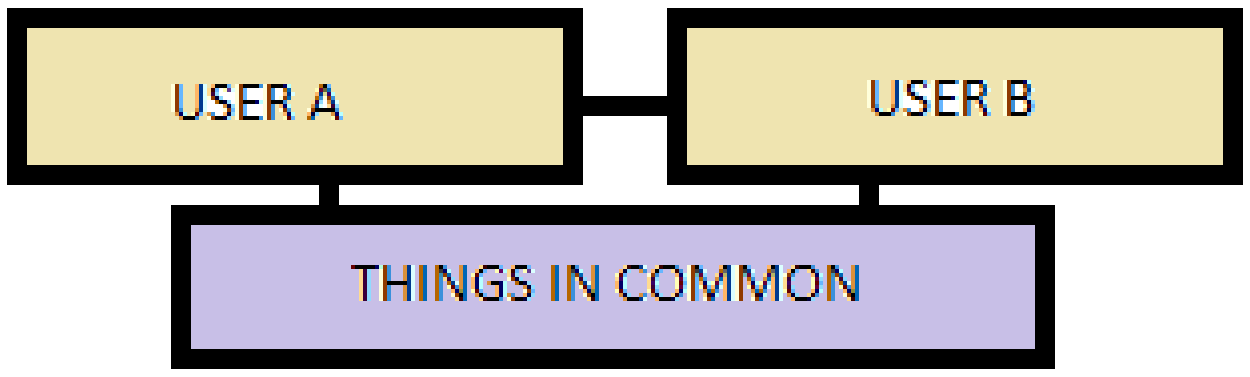

**Figure 9 Common Attributes of User A and User B**

4. PROFILE SIMILARITY

**4.1 Profile Similarity**

On the basis of profile similarity which is our interest of study, the similarity of profiles shows the interaction and closeness levels by measuring the profile information provided by the user. In this proposed work, similarity is measure between a user profile and his friends that are directly linked/ hyperlinked from his/her homepage. In order to make a fair comparison between the profiles, we can equalize the total number of matches made by introducing threshold similarity value for which we would declare a match. So we calculate similarity score and use the decision making algorithm for deciding the match between a source profile and his/her friend's network. The below figure 10, depicts the proposed architectural components.

We evaluate the performance of the profile matching technique by computing the similarity score using string similarity function for all the users with respect to a single individual. In this way, we can predict whether one person is associated with another, by providing a score to all users by their similarity, as friends tend be more similar. Hence the more things two people have in common, the more likely they are to be friends, and the more likely they are to link each other on their Facebook homepages.

We expect users linking to each other on their Facebook homepages to be more similar other than randomly chosen pairs. We measure this effect by a step by step process.

**Step1:**

First we measure how many individuals have common in them and are linked to each other and assign a binary weight to each attribute of the profile with respect to the individual home page. i.e we assign a weight for the number of items in common by comparing attributes and matching each profile's information. The weight increases as per the match of each attribute. It follows the algorithm in assigning the weights to attributes.

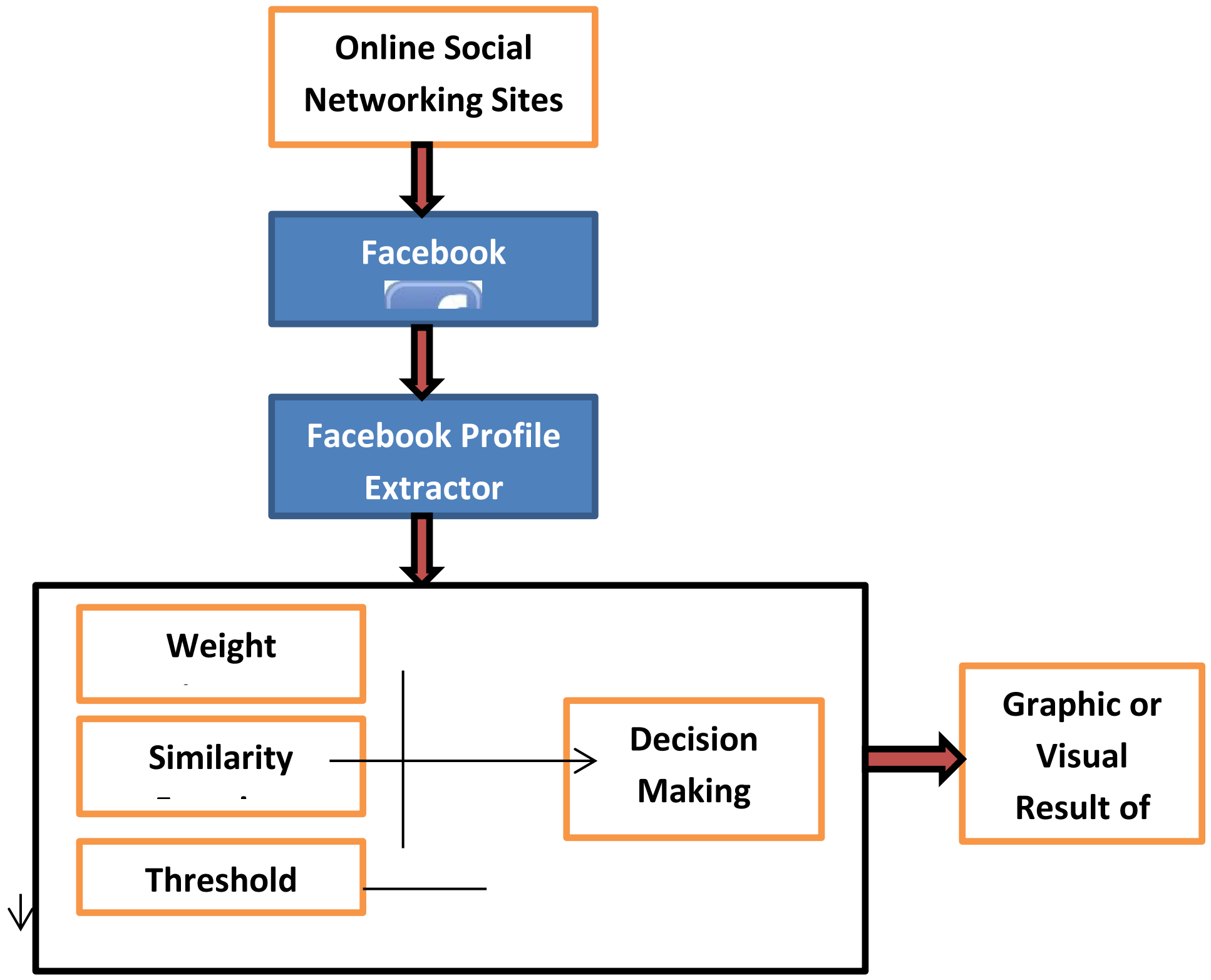

**Figure 10  Architectural Components**

**Step 2:**

After step1 is completed, similarity function assignment is used. We use any of the similarity function assignment techniques which were mentioned earlier. A similarity score is calculated for the user profile. A new similarity score is calculated based on the similarity function used. A threshold value can be estimated by taking the average of the similarities scores.

**Step 3:**

In step 3, a decision making algorithm is used to decide whether there are similarities between the user profiles and target profile. It considers the similarity score calculated and threshold value. If the similarity score is greater than the threshold value then there is a similarity otherwise no.

Finally, one may expect that friends should have the most common, while friends of friends (mutual friends ) should have less in common.

**4.2.1 Attribute Weight Assignment:**

This component mainly aims to assign a weight to each attribute to the user profile. If two profile users mention the same city or working in same organization it shows strong connection than the users with different cities or organizations.An algorithm is proposed to decide weight of different fields[2].

**Attribute Weight Assignment Algorithm:**

**Input**: P: Set of User Profiles

A: Set of attributes used to describe profiles

**Output**: w: Vector of weights assigned to attributes

1. Consider all the user profiles as input to be compared with the source profile.
2. Consider the user profile attributes as ‘a’ to which weights are assigned.If there is a match between the attributes of source profile and user profile then weight of the attribute is assigned as 1,otherwise 0.
3. Begin

```
        foreach Pi in P do
                foreach Pj in P\Pi do

                        a in A

                if( Pi.a==Pj.a) then
                foreach ai in (Pi Pj) do

                        w[a]=1
                        else
                        w[a]=0
                count++
                end
                end
                end
                return w
                end
```

**Binary Weight Assignment:**

According to the algorithm, binary weights are assigned to every field.In this technique, binary weights ( 0 or 1) is assigned to every fields or attributes of the user profiles.

**Table 3 Binary Weight assignment**

| Friends List | Hometown | City | Occupation | Gender | Pages Liked | Total Weight |
|---|---|---|---|---|---|---|
| F1 | 0 or 1 | 0 or 1 | 0 or 1 | 0 or 1 | 0 or 1 | W(F1) |
| F2 | 0 or 1 | 0 or 1 | 0 or 1 | 0 or 1 | 0 or 1 | W(F2) |
| Fn | 0 or 1 | 0 or 1 | 0 or 1 | 0 or 1 | 0 or 1 | W(Fn) |

The above table describes the weight assignement to each attribute of the user profiles.

**Weight of Mutual Friends and Mutual Communities:**

(1) Mutual Friends : The mutual friends specify the mutual social connection between the users. When two or more friends having greater number of mutual friends then they create mutual social network. Weights of mutual friends are added in resultant weights [1].

$$Mfi = \frac{\frac{Mutual\ Friends(U, fi)}{Total\ No.of\ U's\ friends} * 100}{WAF}$$

Where,U=Target Profile

$f_i$= friends profiles

WAF= weight adjustment factor which decides upper limit of weight.

(2) Mutual Communities : The mutual communities or groups specify the mutual interest between users. When two or more friends have greater number of mutaul communities or groups then they are more closed according to shared interest.So, weights are added to the resultant weight [1].

$$MCi = \frac{\frac{Mutual\ Communities(U, fi)}{Total\ No.of\ U's\ Communities} * 100}{WAF}$$

Where,U=Target Profile

$f_i$= friends profiles

WAF= weight adjustment factor which decides upper limit of weight.

### 4.1.2 String Similarity Metrics

When two profiles are compared, their profile attribute fields are compared. There are various ways to measure the similarity score between two textual or string values and can be grouped into 2 main categories.

(1) Syntactic- based similarity approaches
(2) Sematic- based similarity approaches

Syntactic- based similarity approaches provide approximate lexicographical matching of two vales. The approximate string matching techniques can be used to compute the distance between two values that have a limited number of different characters [2].

Semantic- based similarity approaches are used to measure how two values, lexicographically different, are semantically similar. They can be Knowledge- based and Corpus-based.

(1) **Jaro Metric Similarity:** Jaro metric is considered as one of the optimal measures to be primarily intended for comparison. It is based on the number and order of the common characters between two strings [12].
The jaro distance similarity between the two strings s and t can be computed as:

$$Sim_{Jaro}\ ( s, t\ ) = \frac{1}{3}\left(\frac{|s'|}{|s|} + \frac{|t'|}{|t|} + \frac{|s'| -\ 0.5 * T_{s',t'}}{s'}\right)$$

where:
|s| and |t| are the length of each string,
|s'| and |t'| are the number of common characters,
T is number of transposed characters.

(2) **Cosine Similarity**:
Cosine similarity is the best metric which is used frequently when trying to determine similarity between two texts. By determining the cosine similarity, the user is effectively trying to find cosine of the angle between the two objects. For cosine similarities resulting in a value of 0, the documents do not share any attributes (or words) because the angle between the objects is 90 degrees [12].

$$\cos\left(\overrightarrow{t1},\quad \overrightarrow{t2}\right) = \frac{\overrightarrow{t1}\,.\,\overrightarrow{t2}}{||\overrightarrow{t1}||\,||\overrightarrow{t2}||}$$

Where, t1 and t2 vectors representing the vectors of profile attributes.

In understanding the similarity, cosine similarity captures the scale invariant. A stronger property in cosine similarity is that it does not depend on length. Cos ($\alpha\overrightarrow{t1}$, $\overrightarrow{t2}$) = cos ($\overrightarrow{t1}$, $\overrightarrow{t2}$) for $\alpha > 0$. This popularity of the cosine similarity is it allows documents with the same composition but different tools are treated identically.

**Numeric-based attributes**: The Edit distance metric is the most suited technique to compute similarity for this kind of attributes. By calculating cost of minimum number of editing operations called edit script that converts s to t, to measures the distance between two strings, s and t. The edit distance similarity between two values s and t can be evaluated as [12]:

$$Sim_{edit\ distance}(s, t) = 1 - \frac{d}{\max(ls, lt)}$$

where
s and t= comparison values

d= distance between s and t

ls and lt = length of s and t

max (ls, lt) = maximum length between s and t.

In the following, we define :

(1) Computing new similarity scores between profiles using the old similarity score functions

(2) Computing the profile threshold matching:

### 4.1.3 Computing Similarity score between two profiles:

To check with the similarity score, the values of common attributes in both profiles are extracted and their similarity scores are computed.The obtained similarty score are tuned in order to have more realistic score that take into consideration the importance assigned to each attribute. By this way, the new similarity value will tend to increase or decrease depending on the importance of each attribute. A new output generated is the new similarity score to each attribute by applying a weight to the computed similarity scores. The new similarity score is as follows:

$$\boldsymbol{Sim'(P1.a_i, P2.a_i) = \frac{2 * Sim(P1.a_i, P2.a_i) * W(F)}{1 + (\,Sim(P1.a_i, P2.a_i) * W(F))}} \in [\,\mathbf{0}, \mathbf{1}\,]$$

where:

$a_i$ = an attribute used to describe a profile,
P1.$a_i$ and P2.$a_i$ = two values of an attribute $a_i$ in Profile P1 and P2,
W(F)= The total assigned weight of an attribute to the user profile ϵ [0,1],
sim (P1.$a_i$,P2.$a_i$)= similarity score computed between values of an attribute in P1 and P2 ϵ [0,1],
sim'(P1.$a_i$,P2.$a_i$)= the new similarity score computed between the values of an attribute in P1 and P2 ϵ [0,1].

### 4.2.4 Computing the profile threshold matching:

It is the minimal similarity value required for matching two profiles. To compute the threshold value, we use the weights assigned to each attribute. Based on this, the weights form reliable measures and can be considered as reference values for computing a profile matching threshold. It can be evaluated as:

$$T = \frac{Sim'(P1.a_i, P2.a_i) + Sim'(P1.a_i, P3.a_i) + \cdots Sim'(P1.a_i, Pn.a_i)}{Total\ number\ of\ similarity\ score}$$

where T= the threshold to compute,

a= attributes used to describe user profile,

The new similairty scores are sent to the decision making algorithm.The decision making algorithm will return a value, as match or no match between the profiles.

### 4.2.5 Decision Making Algorithm:

The decision making algorithm invloves few steps to decide whether the target profile is similar to the user profiles.If yes, it will return a match otherwise no match [2].

**Input**: Profiles P1 and P2 of two users

P1.a and P2.ai are two values of an attribute a in P1 and P2
**Output**: Result :Similar / Not Similar
Begin
Foreach ai in (P1 P2) do
D=Total Simlarity Score
T= Threshold Value
If D ≥ T then
Result=Similar
Else
Result= Not similar.
End
Return Result
End.

Our match-making technique is based on the well- established result in sociology that friends tend to be similar. Two individuals who are friends can falsely appear to have nothing common if one or both have very little information on their Facebook homepages. It can also happen if the users use their homepages to express different interests. They might both share an interest in books and sports, but one might devote entirely his/ her Facebook homepage entirely to books, while others devotes it only to sports. In this case we would not be able to assign a similarity score with respect to each other, hence we check the personal information such as studies, university etc. and ranks them because there would be no overlap.

Consider a network of five friends connected with the target user profile. Consider the fields of the Facebook given by all the users as the attributes for the algorithm. Each user is connected to the target profile user and can be represented as a graph where each node denotes the users and the edge denotes a link connecting the users.

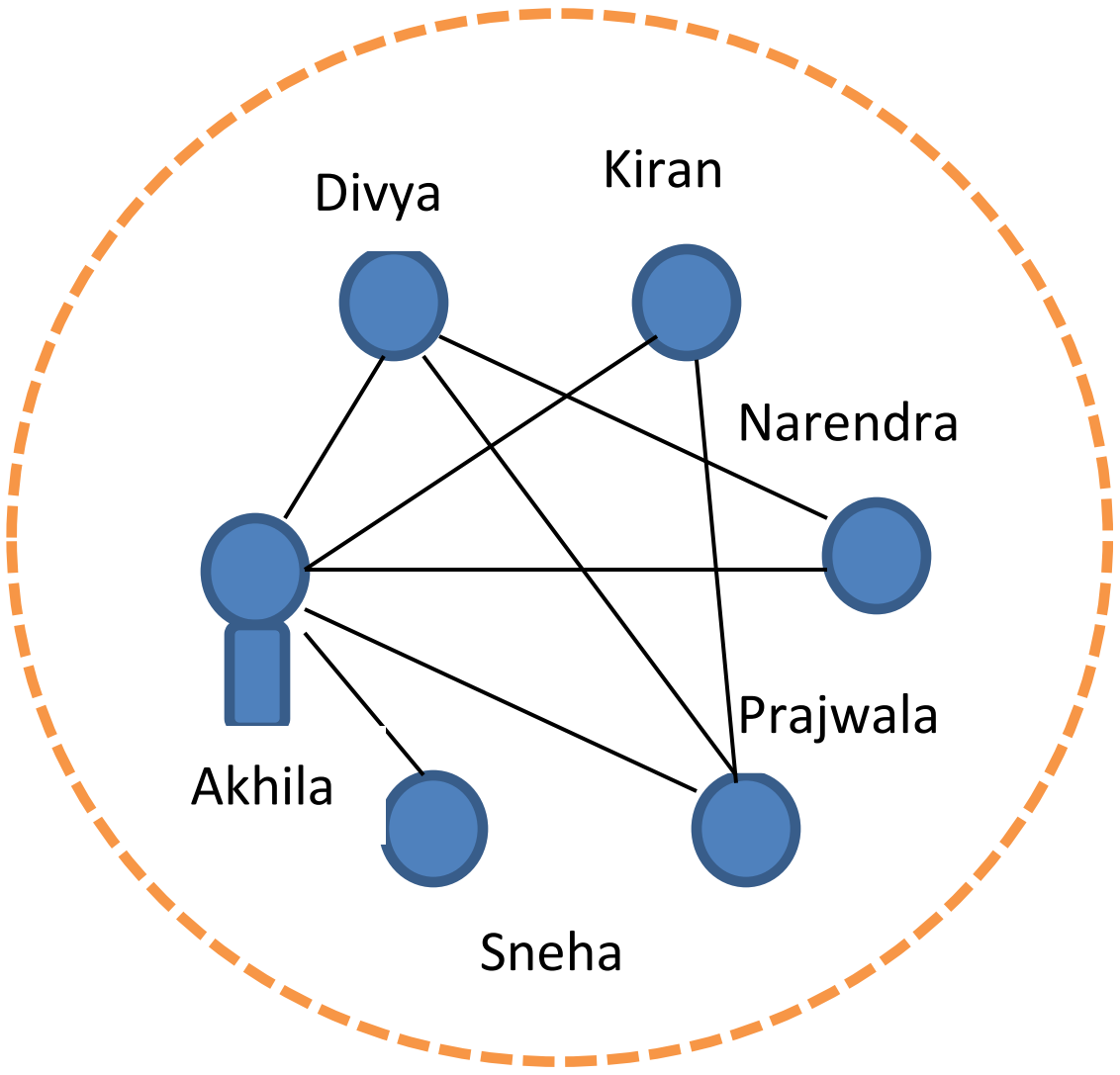

**Figure 11 Facebook Friends Network**

The above figure represents a network of five friends connected in Facebook social network. Each node is a Facebook user and the edge connecting are the relationships with the user profile.

Each node is internally linked to other node, i.e. the target user profile contains mutual friends connecting to each other.

The similarity score of the profiles can be calculated and the similarity of profiles can be extracted by the step by step procedure as mentioned earlier. As the first step, the weights are assigned to the attributes that are considered from the Facebook profile.

For the above network graph, let us consider the attributes.

**Table 4 Binary Weight Assignment for Example Network**

| Friends List | Home town w(H) | City w(C) | Education w(E) | Occupation w(O) | Languages w(L) | Gender w(G) | Sports w(S) | MF w(MF) | MC w(MC) | W |
|---|---|---|---|---|---|---|---|---|---|---|
| Akhila | 1 | 1 | 1 | 1 | 1 | 1 | 1 | 1 | 1 | 9 |
| Divya | 1 | 1 | 1 | 1 | 0 | 1 | 0 | 1 | 1 | 7 |
| Kiran | 1 | 0 | 1 | 0 | 1 | 0 | 1 | 1 | 1 | 6 |
| Narendra | 0 | 0 | 1 | 0 | 0 | 0 | 1 | 1 | 1 | 4 |
| Prajwala | 0 | 0 | 0 | 0 | 1 | 1 | 0 | 1 | 1 | 4 |
| Sneha | | 1 | 0 | 0 | 0 | 1 | 1 | 1 | 0 | 5 |

Table 4 shows the binary weights assigned to the attributes for all the users in the profile. If the attributes match with the target user profile then, the weight assigned is 1 otherwise 0.The total weights of the attributes are calculated by

$$W= \sum_{i=0}^{n} w(H) + w(C) + w(E) + w(O) + w(L) + w(G) + w(S) + w(MF) + w(MC)$$

Now, Similarity function assignment can be used to calculate the similarity scores in profiles. By using cosine similarity function, the similarity score can be calculated.

$$\cos(A, D) = \frac{1.1+1.1+1.1+1.1+1.0+1.1++1.0+1.1+1.1}{\sqrt{1^2+1^2+1^2+1^2+1^2+1^2+1^2+1^2+1^2} . \sqrt{1^2+1^2+1^2+1^2+1^2+1^2+1^2+1^2+1^2}} = 0.88$$

$$\cos(A, K) = \frac{1.1+1.0+1.1+1.0+1.1+1.0++1.1+1.1+1.1}{\sqrt{1^2+1^2+1^2+1^2+1^2+1^2+1^2+1^2+1^2} . \sqrt{1^2+1^2+1^2+1^2+1^2+1^2}} = 0.83$$

Similarly, the cosine similarity metric can be used to find the cosine values of all the profiles in the Facebook network. Here, A represents profile P1 and D represents profile P2 i.e. A represents Akhila and D represents Divya. A tabular form can be constructed to estimate the cosine values of all the profiles in the network.

**Table 5 Cosine Similarity Scores**

| Friends Lists | Cosine Similarity [Sim( P1, P2)] |
|---|---|
| Divya (D) | 0.88 |
| Kiran (K) | 0.83 |
| Narendra (N) | 0.66 |
| Prajwala (P) | 0.66 |
| Sneha (S) | 0.74 |

Table 5 represents the cosine values of all the profiles considering profile attributes as vectors for cosine similarity metric. Now, let us consider the new similarity score that is introduced earlier, to calculate the new similarity score using the cosine values in the table and it is computed as follows:

$$\boldsymbol{Sim'(P1.a_i, P2.a_i) = \frac{2 * Sim(P1.a_i, P2.a_i) * w(F)}{1 + (Sim(P1.a_i, P2.a_i) * w(F))}} \in [\, \mathbf{0}, \mathbf{1} \,]$$

For the profile in the given network,

$$Sim'(A, D) = \frac{2 * Sim(A, D) * 7}{1 + (Sim(A, D) * 7)} \in [\, 0,1 \,]$$

The new similarity score of *Sim ‘ (A, D) = 1.73*.The same follows for all the profiles and the new similarity score is calculated. A tabular form can be constructed with all the new similarity scores and similar profiles can be estimated by the decision making algorithm.

**Table 6 New Similarity Scores**

| Friends List (F) | New Similarity Scores |
|---|---|
| Divya | 1.73 |
| Kiran | 1.66 |
| Narendra | 1.45 |
| Prajwala | 1.45 |
| Sneha | 1.50 |

Table represents the new similarity scores calculated by the formula using the cosine similarity function. A threshold value can be calculated by considering the average of all the new similarity scores.This threshold value is used to estimate the most similar profile by the decision making algorithm.

$$T = \frac{Sim'(P1.a_i, P2.a_i) + Sim'(P1.a_i, P3.a_i) + \cdots Sim'(P1.a_i, Pn.a_i)}{Total\ number\ of\ similarity\ score}$$

The threshold value can be calculated for the profiles in the network by considering the average of all the similarity scores.

$$T = \frac{1.73 + 1.66 + 1.45 + 1.45 + 1.50}{5}$$

The similarity threshold value, T= 1.57. The similarity of the profiles can be determined by the final step of checking the simlarity score is greater than or equal to the similarity threshold value calculated. Step 3, metioned earlier, considers the decision making algorithm which takes the similarity threshold value and new similarity scores to decide whether the profile is similar or not. By evaluating, the threshold and the similarity scores, the results shows that Divya and Kiran are the most similar profiles as the similarity scores are greater than or equal to the threshold value(1.57). The result can be seen in a plotted graph where the new similarity score are considered.

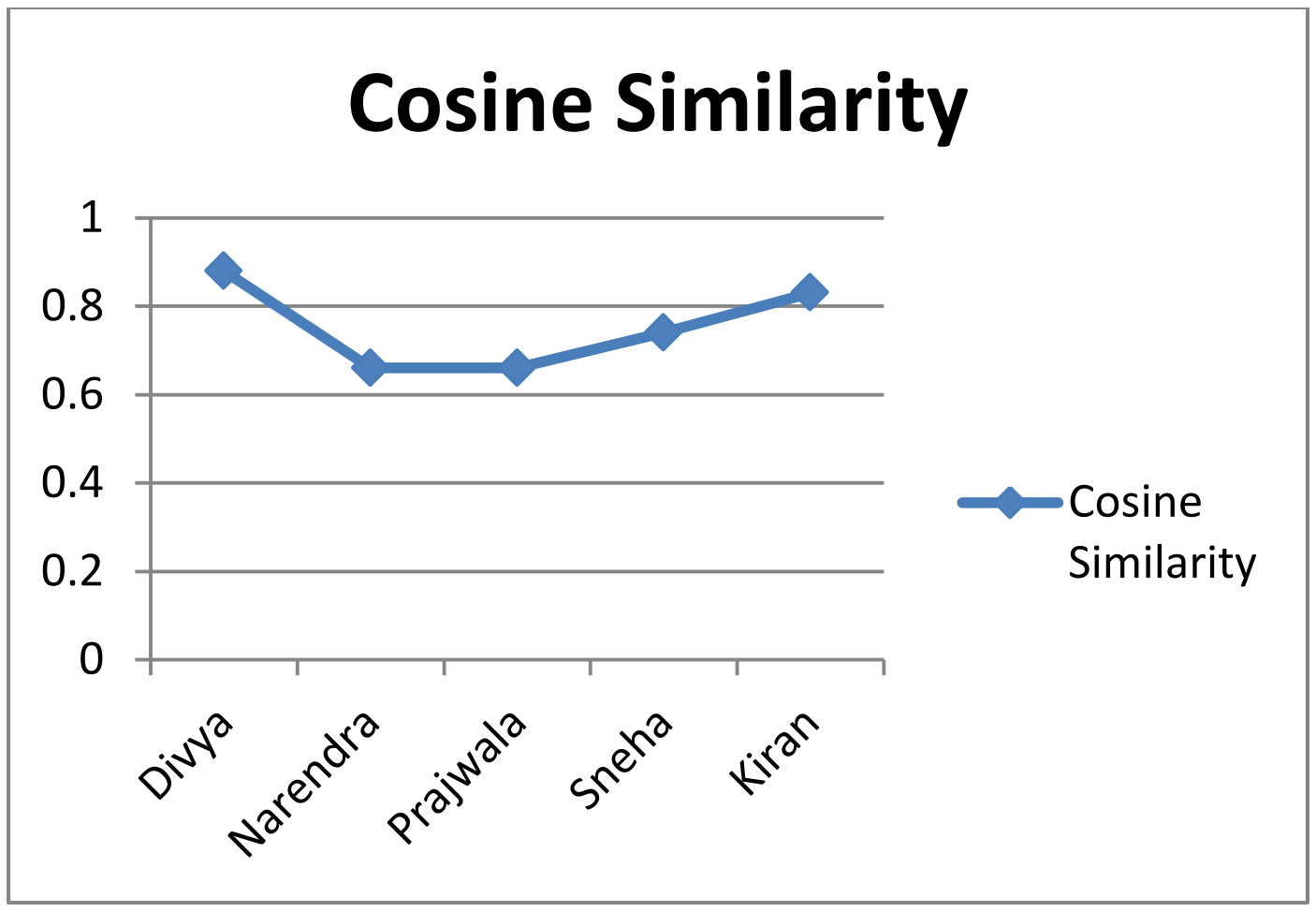

**Figure 12 Cosine Similarity Graph**

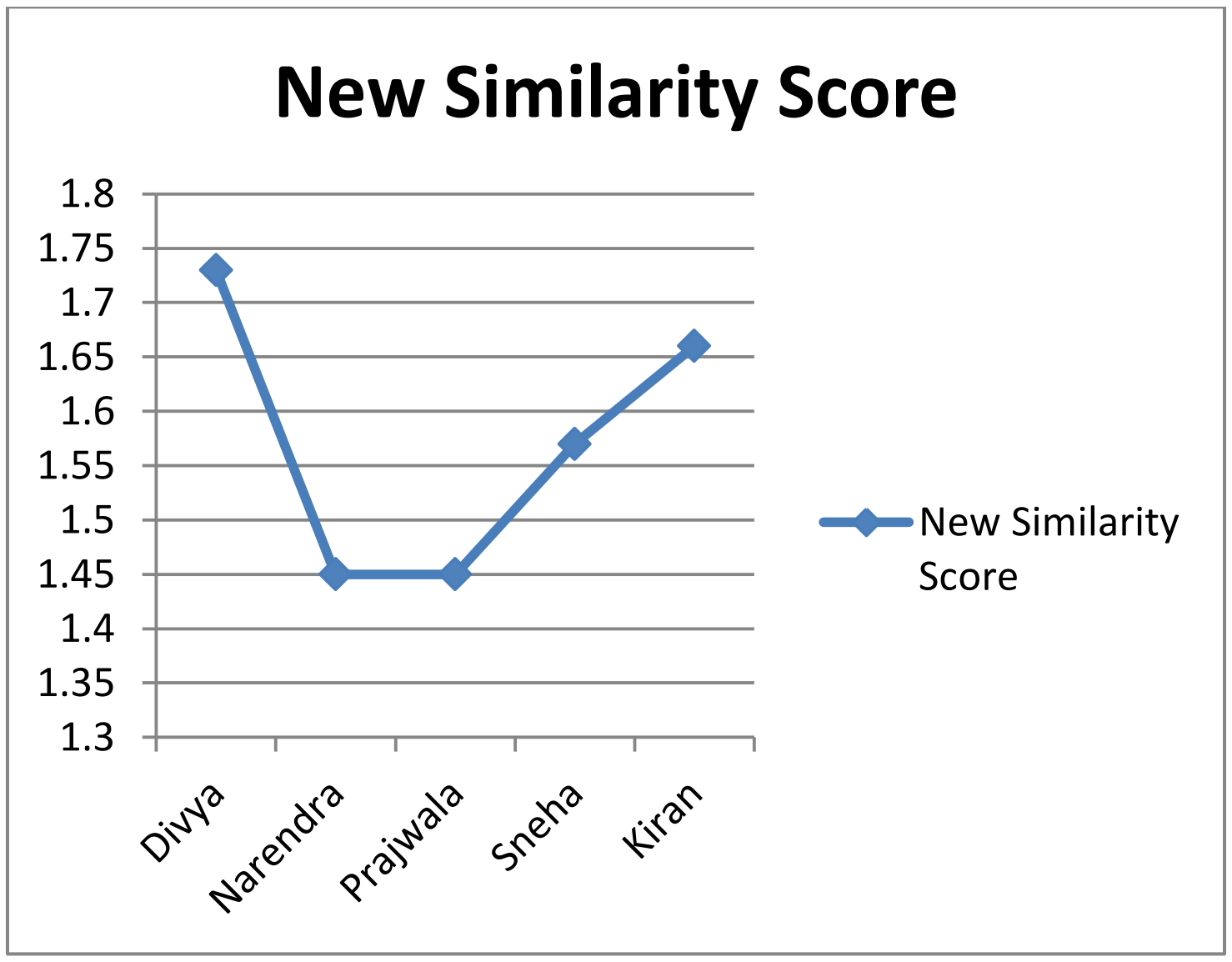

**Figure 13 New Similarity Score Graph**

The above figures represents the cosine similarity and new similarity function graphs used to calculate similar profiles. The graph clearly states that Divya and Kiran, having the similarity score greater than the threshold value are most similar to the target user profile. The results clearly states that the proposed work is able to discover the biggest possible number of profiles that are similar to the target user profile, as the existing techniques are unable to detect. In our work, attributes were assigned weights manually, string and semantic similarity metrics were used to compare attributes values thus predicting the most similar profiles. Other metrics such as Bhattacharyya distance can also be used to determine similarity.

## REFERENCES

[1] Rajni Ranjan Singh,Deepak Singh Tomar, "Approaches for user profile Investigation in Orkut Social Network", Maulana Azad National Institute of Technology (MANIT), Bhopal, India, 2009.

[2] Elie Raad, Richard Chbeir, Albert Dipanda, "User Profile Matching in Social Networks",published in "Network Based Information Systems, Japan 2010.

[3] Robert Patton, "Facebook and Networked Interactivity", December 2007.

[4] Lada A. Adamic∗, Eytan Adar, "Friends and neighbors on the Web", HP Labs, 1501 Page Mill Road, Palo Alto, CA 94304, USA.

[5] Dinithi Pallegedara, Lei Pan, "Investigating Facebook Groups through a Random Graph Model".

[6] Facebook Network analysis using Gephi, www.gephi.com.

[7] William H. Hsu Joseph Lancaster Martin S.R. Paradesi Tim Weninger, "Structural Link Analysis from User Profiles and Friends Networks: A Feature Construction Approach", Department of Computing and Information Sciences, Kansas State University.

[8] Naohiro Matsumura, David E. Goldberg, Xavier Llora, "Mining Directed Social Network", Message Board.

[9] Patrick Doreian, Tom A.B. Snijders, "Social Networks, an international journal of structural analysis".

[10] Minas Gjoka, Maciej Kurant, Carter T. Butts, Athina Markopoulou.," A walk in Facebook: Uniform Sampling of Users in Online Social Networks".

[11] Graph Visualization of an Economic Environment using Gephi.

[12] Alexander Strehl, Joydeep Ghosh, and Raymond Mooney, "Impact of similarity measures on Web-page Clustering", The University of Texas, Austin, Texas.

[13] S. Kak, Feedback neural networks: new characteristics and a generalization. Circuits, Systems, and Signal Processing, vol. 12, pp. 263-278, 1993.

[14] S. Kak, Self-indexing of neural memories, Physics Letters A, vol. 143, pp. 293-296, 1990.

[15] S. Kak and M.C. Stinson, A bicameral neural network where information can be indexed. Electronics Letters, vol. 25, pp. 203-205, 1989.

[16] D.L. Prados and S. Kak, Neural network capacity using the delta rule. Electronics Letters, vol. 25, pp. 197-199, 1989.

[17] D. J. Watts and S. H. Strogatz, "Collective dynamics of 'small-world' networks.," Nature, vol. 393, no. 6684, pp. 440–2, Jun. 1998.

[18] J. Kleinberg, “The small-world phenomenon : an algorithmic perspective,” in Proceedings of the thirty-second annual ACM symposium on Theory of computing - STOC ’00, 2000, pp. 163–170.
[19] H. Ebel, L.-I. Mielsch, and S. Bornholdt, “Scale-free topology of e-mail networks,” Phys. Rev. E, vol. 66, no. 3, Sep. 2002.
[20] D. J. Watts, “The ‘New’ Science of Networks,” Annu. Rev. Sociol., vol. 30, no. 1, pp. 243–270, Aug. 2004.
[21] Facebook Data Team, “Anatomy of Facebook,” 2012. [Online]. Available: https://www.facebook.com/notes/facebook-data-team/anatomy-of-facebook/10150388519243859 . [Accessed: 06-Jan-2012].
[22] S. Kak, On generalization by neural networks. Information Sciences, vol. 111, pp. 293-302, 1998.
[23] K.-W. Tang and S. Kak, A new corner classification approach to neural network training. Circuits, Systems, and Signal Processing, vol. 17, pp. 459-469, 1998.
[24] S. Kak, New algorithms for training feedforward neural networks. Pattern Recognition Letters, vol. 15, pp. 295-298, 1994.
[25] S. Kak, On training feedforward neural networks. Pramana, vol. 40, pp. 35-42, 1993.
[26] S. Kak, Faster web search and prediction using instantaneously trained neural networks. IEEE Intelligent Systems, vol. 14, pp. 79-82, November/December 1999.
[27] S. Kak, A class of instantaneously trained neural networks. Information Sciences, vol. 148, pp. 97-102, 2002.
[28] K.W. Tang and S. Kak, Fast classification networks for signal processing. Circuits, Systems, Signal Processing, vol. 21, pp. 207-224, 2002.
[29] S. Asur and B. A. Huberman, “Predicting the future with social media,” in 2010 IEEE/WIC/ACM International Conference on Web Intelligence and Intelligent Agent Technology, 2010, pp. 492–499.
[30] A. Tumasjan, T. O. Sprenger, P. G. Sandner, and I. M. Welpe, “Predicting elections with Twitter : what 140 characters reveal about political sentiment,” in Proceedings of the Fourth International AAAI Conference on Weblogs and Social Media, 2010, pp. 178–185.
[31] S. Yu and S. Kak, “A Survey of Prediction Using Social Media,” 07-Mar-2012. [Online]. Available: http://arxiv.org/abs/1203.1647. [Accessed: 11-Mar-2013].
[32] Michael H Goldhaber, “The Attention Economy and the Net,” First Monday, vol. 2, no. 4, pp. 1–27, 1997.
[33] N. Pope, “The Economics of Attention: Style and Substance in the Age of Information (review),” Technol. Cult., vol. 48, no. 3, pp. 673–675, 2007.
[34] S. Yu and S. Kak, “Social Network Dynamics: An Attention Economics Perspective,” in Social Networks: A Framework of Computational Intelligence, Edited by Witold Pedrycz and Shyi-Ming Chen. Springer, 2014.
[35] H. A. Simon, “Designing organizations for an information rich world,” in Computers, communications, and the public interest, 1971.
[36] S. Yu and S. Kak, An empirical study on how users adopt famous entities. International Conference on Future Generation Communication Technologies, London, December 2012.